\documentclass[12pt]{iopart}
\usepackage{graphicx}
\usepackage{epsfig}
\usepackage{color}

\begin{document}

\title[Neutron star equation of state]{Neutron star equation of state
and GW170817}

\author{Sajad A. Bhat and Debades Bandyopadhyay}

\address{Astroparticle Physics and Cosmology Division, Saha Institute of 
Nuclear Physics, 1/AF Bidhannagar, Kolkata-700064, India and
Homi Bhabha National Institute, Training School Complex,
Anushaktinagar, Mumbai-400094}
\begin{abstract}
Properties of neutron stars in GW170817 are investigated using different 
equations of state (EoSs) involving nucleons, $\Lambda$ hyperons, quarks 
resulting in $2M_{\odot}$ neutron stars. This calculation is performed using 
the same EoS for merger components and for low spin prior case.
It is found from the computations of tidal deformability parameters that soft 
to moderately stiff equations of state are allowed by the 50$\%$ and  
90$\%$ credible regions obtained from the gravitational wave observation of
binary neutron star merger GW170817, whereas the stiffest hadron-quark EoS 
which lies above the upper 90$\%$ limit, is ruled out. A correlation 
among the tidal deformabilities and masses is found to exist as already
predicted. Furthermore moments of inertia and quadrupole moments of merger 
components of GW170817 are estimated. 
\end{abstract}

\noindent{\it Neutron stars, dense matter, equations of state, gravitational
waves}

\submitto{\jpg}
\maketitle

\section{Introduction}
The discovery of gravitational waves from the binary neutron star merger 
event GW170817 followed by the detection of its transient counterparts across 
the electromagnetic (EM) spectrum has heralded a new era in multimessenger 
astrophysics\cite{abbott1,abbott2}. The observed short Gamma Ray Burst (GRB) 
1.7 s after the coalescence time provides clinching evidences for the
association of short GRBs with neutron star mergers. This discovery of two 
colliding neutron stars and its aftermath led to plethora
of information about short Gamma Ray Bursts, binary chirp mass, tidal 
deformability and dense matter in neutron star interior, speed of 
gravitational waves and heavy element synthesis due to r-process in ejected 
neutron-rich matter. The gravitational wave data led to the estimation of 
the binary chirp mass, 
${{\cal M}_{chirp}} = ({m_1} {m_2})^{3/5}/({m_1} + {m_2})^{1/5}$ in the
90$\%$ credible interval as 1.188$^{+0.004}_{-0.002}$ M$_{\odot}$ 
\cite{abbott1}. For low 
spin prior, the component masses of the binary lie in the range 
1.17-1.6 M$_{\odot}$, whereas the total mass of the binary is 
2.74$^{+0.04}_{-0.01}$ M$_{\odot}$. The binary mass ratio ($q = m_2/m_1$) is 
constrained in the range 0.7 - 1.0 for low spin prior.  
The optical/infrared transient, several hours after GW170817, was found to be
consistent with emissions of a Kilonova which shines through radioactive
decays of r-process nuclei synthesised in the neutron-rich ejected matter
\cite{metzger,metzger10}.

The observation of GW170817 reveals many interesting aspects of dense matter
in neutron stars and its equation of state (EoS). The fate of the compact 
remnant formed in the binary neutron star merger might be closely related
to the amount of ejected material as estimated from EM signals \cite{margalit}.
A prompt collapse is ruled out by the quantity of blue kilonova ejecta observed
in optical wavelengths. It is argued that the merger remnant was born as a 
hypermassive neutron star (HMNS) supported by differential rotation
for a short duration of time. This picture of short lived HMNS might be
consistent with the large quantity of red Kilonova ejecta, as observed in the
infrared, originating from the accretion torus around the HMNS before 
its collapse to a black hole. The compact remnant spined down emitting 
gravitational waves and might have collapsed to a black hole close to the 
mass-shedding limit of uniformly rotating neutron stars \cite{rezzolla}. This
conclusion about the merger remnant leads to the upper
bound on the maximum mass of non-rotating neutron stars and much tighter
constraint on the EoS of dense matter \cite{margalit,rezzolla,radice,shapiro}. 
The lower bound on the maximum mass of neutron stars is obtained from pulsar 
observations. 

It was long debated that the tidal effects in the late inspiralling phase of 
binary neutron stars could be large and detected by gravitational wave 
detectors \cite{flan,hind,read}. The tidal deformation of a neutron star might 
provide crucial information about the dense matter EoS. The effective tidal
deformability parameter is expected to be 
determined from gravitational wave signals. Indeed this was achieved in
GW170817. The LIGO and VIRGO observations of GW170817 placed an upper limit 
on the dimensionless combined tidal deformability ${\bar{\Lambda}} \leq 800$ in
the low spin case at 90$\%$ confidence level \cite{abbott1}. A lower limit on
$\bar{\Lambda} \geq 400$ was estimated from the observational data of the 
electromagnetic counterpart of GW170817 combined with numerical relativity 
simulations \cite{radice}. Recently another alternative approach involving
radiative transfer simulations for the electromagnetic transient AT2017gfo
predicts the lower bound on the tidal deformability to be 
$\bar{\Lambda}\geq 197$ \cite{coughlin}.  

Gravitational wave data of GW170817 were reanalysed by De et al. \cite{de}.
The initial analysis of the LIGO/VIRGO collaboration (LVC) differed from that 
of De et 
al. in the sense that the same EoS was not used for two neutron stars of 
GW170817 in the former case whereas in the latter case the tidal 
deformabilities ($\Lambda_1$ and $\Lambda_2$) and masses of both neutron stars 
were connected through the relation 
${\Lambda_1}/{\Lambda_2}\sim q^6$ implying that both neutron stars are 
described by the same EoS. Later the LVC analysed the data again using 
correlations in tidal deformabilities \cite{ligo2}. The correlation among tidal 
deformabilities led to a 20$\%$ reduction in 90$\%$ confidence upper 
bound of the effective tidal deformability \cite{de,ligo2}. Tighter bounds on
${\Lambda_1}/{\Lambda_2}$ tuned to chirp mass are prescribed in an 
EoS-independent manner for gravitational waveform analysis \cite{zhao}.      

The lower and upper bounds on the tidal deformability parameter put strong 
constrains on the dense matter EoS in neutron star interior. Radice et al. 
showed that too soft or too stiff EoSs were rejected because of those 
constraints \cite{radice}. Masses, radii and moments of inertia of neutron 
stars are the direct probes of dense matter EoS. The knowledge of tidal 
deformability from GW170817 was exploited to constrain the neutron star radius
\cite{de,fattoyev,sch18,ozel,ligo3}.
It was shown that the upper limit of the radius of a 1.4 M$_{\odot}$ neutron 
star was $\leq 13.76$ km \cite{fattoyev}. In another investigation with one
million different EoSs, the radius (R) of a 1.4 M$_{\odot}$ neutron star is 
found to be $12.00 \leq R/km \leq 13.45$ \cite{sch18}. Similar conclusion was
drawn about the radius in Ref.\cite{ozel}. Using the correlation among tidal
deformabilities of merger components, radii of both neutron stars were 
constrained to be $8.7 < R/km < 14.1$ at the 90$\%$ credible interval
\cite{de}. The LVC also estimated the neutron star radii firstly adopting 
EoS-insensitive relations and secondly the same parameterized EoS for both 
neutron stars \cite{ligo3}. In the second case, the condition that the EoS was
compatible with 1.97 M$_{\odot}$ neutron stars was imposed. This led to higher
radii of neutron stars in the second case than those of the first case
\cite{ligo3}. So far we have 
noticed that gravitational wave data from GW170817 as well as its EM 
counterpart AT2017gfo led to the determination of upper bounds on the mass and 
radius of non-rotating neutron stars. Besides masses and radii of neutron
stars, the measured tidal deformability could put constraints on other 
properties of merger components of GW170817 such as moment of inertia and 
quadrupole moment. This motivates us to explore EoS of dense matter
and properties of merger components of GW170817 in this work.        

This paper is organised in the following way. Equations of state used in this 
calculation are described in Sec II. We discuss results in Sec. III. We 
summarise our findings and conclude in Sec. IV.

\section{Formalism}
We discuss the computation of tidal deformability, moment of inertia and 
quadrupole moment in this section. These quantities are EoS dependent 
\cite{deb}. We adopt different relativistic mean field (RMF) models for the 
EoS of beta-equilibrated and charge neutral matter. The strong interaction 
among nucleons from the crust to the core is mediated by the exchange of 
scalar, vector and iso-vector mesons in these RMF models. The RMF 
parameterizations used in this calculations are TM1, TMA, SFHo, SFHx, DD2, 
DDME2 \cite{hem17,ring}. An extended nuclear statistical model developed by 
Hempel
and Schaffner (HS) is used to describe the matter in the sub-saturation density
regime \cite{hs10} along with the RMF nuclear interactions for the nonuniform
and high density matter in all these cases. The SFHo and SFHx EoSs are fitted to
some measurements of neutron star radii \cite{stein13}. The density dependent
(DD) couplings are used in the RMF models of nuclear interactions denoted by 
DDME2 and DD2 
\cite{ring,typ10}. The DD2 RMF model is extended to include $\Lambda$ hyperons 
involving repulsive interaction mediated due to $\phi$ mesons, by Banik, 
Hempel and
Bandyopadhyay (BHB) and is denoted as BHB$\Lambda \phi$ \cite{banik14}. All the
above mentioned EoSs are unified in the sense that the same RMF model used in
the crust and core. Furthermore, these EoS are widely used for core collapse
supernova and neutron star merger simulations. We      
consider also equations of state involving first-order phase transition from 
hadronic matter to quark matter. In one case,  
the hadronic phase including all hyperons and $\Delta$ resonance are described 
by the DD2 RMF model \cite{weber1}. For the quark phase made of u, d and s 
quarks, a non-local extension of the Nambu-Jona-Lasinio model is employed 
\cite{NJL}. Gibbs phase rules are imposed in the hadron-quark (HQ) mixed phase.
We label this EoS as HQ1. The hadronic phase in other HQ EoS made of only
nucleons is calculated using the NL3 RMF model whereas the quark phase is
described by the effective bag model including quark interaction 
\cite{prc91}. The HQ mixed phase in this case is based on the Maxwell 
construction. This EoS is denoted as HQ2.

\begin{figure*}
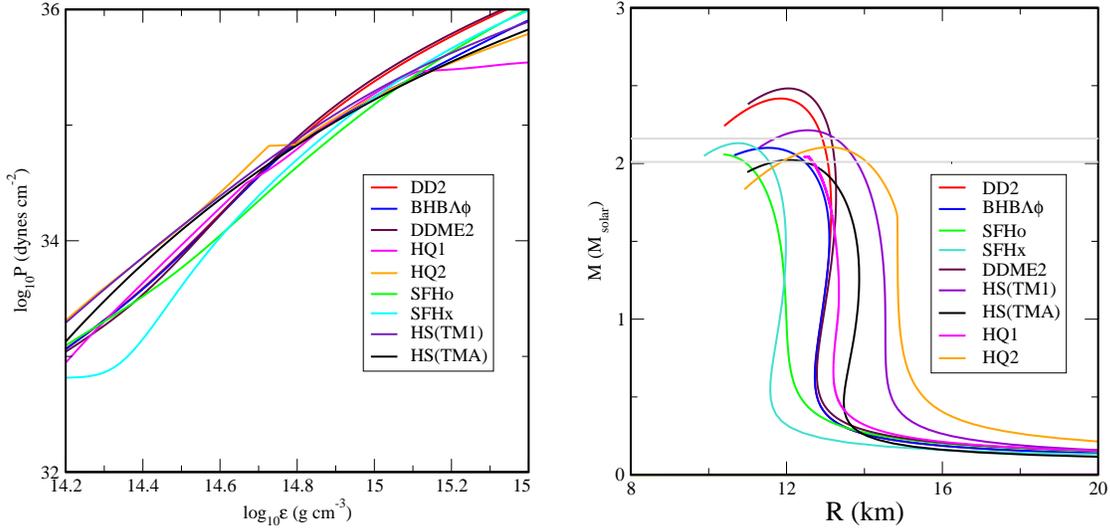

\vspace{0.8cm}
\begin{center}
\resizebox{0.465\textwidth}{!}{
\includegraphics{eos9.eps}
}
\hspace{0.15cm}
\resizebox{0.45\textwidth}{!}{
\includegraphics{mrgw170817.eps}
}
\caption{\small{(Color online)
Pressure versus energy density is shown for different compositions (left panel)
and mass-radius relation is exhibited for those equations of state  
(right panel)}.}
\end{center}
\end{figure*}
A static and spherically symmetric star develops a quadrupole moment ($Q_{ij}$) 
in response to a static external quadrupolar tidal field $\varepsilon_{ij}$. 
The tidal deformability in the linear order is defined as 
$\lambda = -\frac{Q_{ij}}{\epsilon_{ij}}$. The $l=2$ dimensionless tidal Love
number $k_{2}$ related to $\lambda$ is given by \cite{hind}
\begin{eqnarray}
 \lambda = \frac{2}{3}k_{2}R^{5}~.
\end{eqnarray}
The spherically symmetric star under the linear $l=2$, $m = 0$ perturbation 
due to the tidal field results in a static,
even-parity perturbation of the metric which in the Regge-Wheeler gauge 
reduces to \cite{hind} 

\begin{eqnarray}
ds^2 &=& - e^{2\Phi(r)} \left[1 + H(r)
Y_{20}(\theta,\varphi)\right]dt^2
\nonumber\\
& & + e^{2\Lambda(r)} \left[1 - H(r)
Y_{20}(\theta,\varphi)\right]dr^2
\nonumber \\
& & + r^2 \left[1-K(r) Y_{20}(\theta,\varphi)\right] \left( d\theta^2+ \sin^2\theta
d\varphi^2 \right),
\nonumber\\
& &
\end{eqnarray}
where,  $K'(r)=H'(r)+2 H(r) \Phi'(r)$. Finally a second order differential 
equation for metric function $H$ is obtained as   
\begin{eqnarray}
H'' + H' \left(\frac{2}{r}+\Phi '-\Lambda'\right) + H \left(-\frac{6e^{2\Lambda}}{r^2}-2(\Phi ')^2\right.\nonumber\\
\left. +2\Phi'' +\frac{3}{r}\Lambda ' + \frac{7}{r}\Phi '- 2 \Phi'   \Lambda '+ \frac{f}{r}(\Phi '+\Lambda ')\right)=0~,
\end{eqnarray}
where, $f= {d\epsilon}/{dp}$. This equation is integrated outward from the 
center. Applying the asymptotic
behaviour of $H(r)$, the $\ell=2$ tidal love number is given by,
\begin{eqnarray}
k_2 &=& \frac{8C^5}{5}(1-2C)^2[2+2C(y-1)-y]\nonumber\\
      & & \times\bigg\{2C[6-3y+3C(5y-8)]\nonumber\\
      & & +4C^3[13-11y+C(3y-2)+2C^2(1+y)]\nonumber\\
      & & +3(1-2C)^2[2-y+2C(y-1)] \ln(1-2C)\bigg\}^{-1}~,
\end{eqnarray}
$y=RH'(R)/H(R)$ and compactness of the star, $C=M/R$.

The dimensionless tidal deformability, dimensionless moment of inertia and
dimensionless quadrupole moment are defined as $\Lambda_{1,2} = \frac{\lambda_{1,2}}{m^{5}_{1,2}}$,
$\bar{I}_{1,2} = \frac{I_{1,2}}{m^{3}_{1,2}}$ and    $\bar{Q}_{1,2} = \frac{Q_{1,2}}{(m^3_{1,2} (J_{1,2}/m^2_{1,2})^2)}$
respectively, where subscripts 1 and 2 correspond to masses of merger 
components, $m_1$ and $m_2$ respectively. Quadrupole moment $Q_{1,2}$
is compared with the Kerr solution quadrupole moment $J_{1,2}^{2}/m_{1,2}$ and 
the dimensionless $\bar {Q}_{1,2}$ are known as Kerr factors corresponding to 
$m_{1,2}$.
Moment of inertia and quadrupole moment are calculated by the spectral 
scheme within the numerical library LORENE \cite{gourg,lorene}.

\begin{figure}
\vspace{0.7cm}
\begin{center}
\resizebox{0.40\textwidth}{!}{
\includegraphics{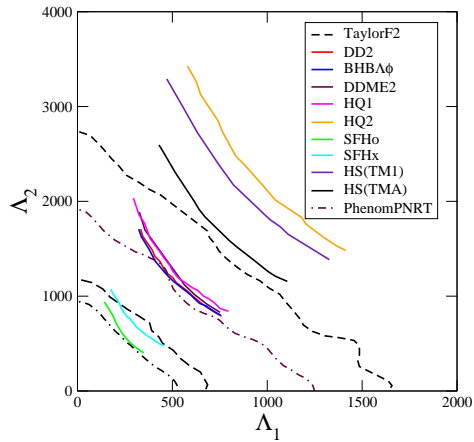}
}
\caption{\small{(Color online)
Dimensionless tidal deformability parameters $\Lambda_1$ and $\Lambda_2$
are plotted here for different equations of state. Dashed and dash-dotted lines 
denote the 50$\%$ and 90 $\%$ probability contours as obtained from 
Ref.\cite{abbott1,ligo2}.}}
\end{center}
\end{figure}

\section{Results and Discussion}
We adopt nuclear EoSs HS(TM1), HS(TMA), HS(SFHo), HS(SFHx), DDME2, HS(DD2), 
hyperon 
EoS BHB$\Lambda \phi$ and hadron-quark EoSs HQ1 and HQ2 for the calculation of 
neutron star properties as described in section II and shown in the left 
panel of Figure 1. The right panel of Figure 1 shows 
neutron star mass as a function of radius for the above mentioned EoSs. All 
those EoSs 
are compatible with 2 M$_{\odot}$ neutron stars. We could learn valuable 
lessons about dense matter EoS from the fate of the massive remnant in GW170817.
It is inferred that a hypermassive neutron star was born in the binary merger
event and later it collapsed to a black hole. In this scenario, different 
groups estimated the upper bound on the maximum mass of non-rotating neutron
stars ($M_{max}^{TOV}$) to be 2.16 M$_{\odot}$ 
\cite{margalit,rezzolla,shapiro,banik17}. On the other hand, the lower limit on 
the neutron star maximum mass 2.01$\pm0.04$ M$_{\odot}$ was obtained from the 
observations of galactic neutron stars. Both bounds on the neutron star maximum
mass i.e. $2.01 \pm 0.04 \leq M_{TOV}/M_{\odot} \leq 2.16 \pm 0.03$, 
put strong constraints on the EoS of dense matter. All EoSs except DD2, 
DDME2 and HS(TM1) satisfy these constraints on M$_{max}^{TOV}$ as shown 
by two horizontal lines in the right panel of Fig. 1.
  
\begin{figure}
\begin{center}
\vspace{0.3cm}
\resizebox{0.45\textwidth}{!}{
\includegraphics{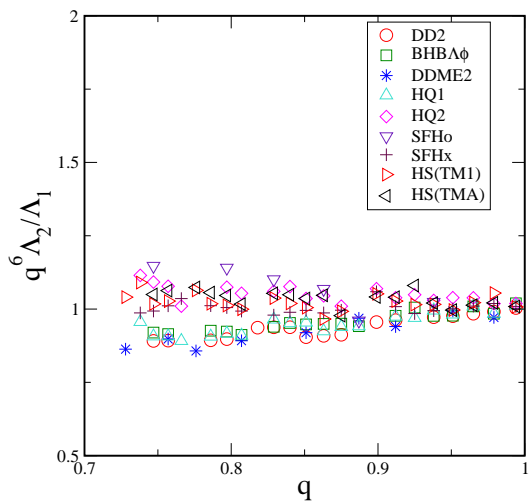}
}
\caption{\small{(Color online)
The quantity $q^6{\Lambda_2}/{\Lambda_1}$ is shown as a function of mass ratio 
($q$) for chirp mass 1.188 M$_{\odot}$ and different equations of state.}}
\end{center}
\end{figure}

Tidal deformability $\Lambda_{2}$ is plotted with $\Lambda_{1}$ in Fig. 2 for 
all those EoSs considered here. Dimensionless tidal deformabilities of both 
merger components are calculated using the same EoS. It also shows the 
50$\%$ and 90$\%$ credible intervals (dashed and dash-dotted lines) 
for the low spin case obtained using waveform models of TaylorF2 and PhenomPNRT 
\cite{abbott1,ligo2}. As the tidal deformability is directly proportional to 
$R^{5}$, the compactness increases from top right corner to bottom left 
corner of Fig. 2. The SFHo EoS represents neutron stars with maximum 
compactness among all EoSs. The HQ2 EoS on the top right corner implies
the least compact neutron stars and lies far outside the 90$\%$ credible 
interval. It is noted that HS(DD2) and BHB$\Lambda\phi$ EoSs which were 
allowed by TaylorF2 model, are now marginally compatible with the 90$\%$ 
contour of PhenomPNRT. The other EoSs which fall well inside 50$\%$ 
and 90$\%$ confidence intervals of PhenomPNRT are validated.
  
It has been noted that the tidal deformability parameter could probe the 
dense matter EoS. This can be further understood from Figs. 1 
and 2. The upper limit on neutron star maximum mass is compatible with HQ2 and 
HS(TMA) along with several other EoSs as evident from Fig. 1. However, HQ2 and 
HS(TMA) EoSs are ruled out by the 90$\%$ confidence contour in Fig. 2. This 
demonstrates that the low density parts of HQ2 and HS(TMA) EoSs are not well 
constrained and lead to larger radii ($>14$ km) for merger components. Besides
this, the nuclear matter EoS in hadron-quark phase transition in HQ2 is 
described by the NL3 EoS which is very stiff. On the other hand, the neutron 
star maximum mass is estimated by the overall EoS which becomes softer due to 
the phase transition to quark matter.   

A correlation among tidal deformabilities and mass ratio of neutron stars
was reported by different groups \cite{de,ligo2,zhao}. When both neutron stars
are described by the same EoS, it is found that tidal deformabilities follow 
the relation ${\Lambda_1}/{\Lambda_2} \sim q^6$ \cite{de,zhao}. Furthermore,
analytical lower and upper bounds on ${\Lambda_1}/{\Lambda_2}$ tuned to chirp
mass were estimated studying large number of piecewise polytropic EoSs with 
and without strong first order hadron-quark phase transitions \cite{zhao}. We
investigate this correlation among tidal deformabilities and mass ratio for
EoSs considered in Figs. 1 and 2. Figure 3 shows $q^6{\Lambda_2}/{\Lambda_1}$
as a function of mass ratio $q$. It is noted that this correlation holds 
good for values of mass ratio $q \geq 0.9$ for most EoSs adopted 
in this work. However, it is observed that the quantity 
$q^6{\Lambda_2}/{\Lambda_1}$ deviates from the value of unity for smaller 
values of $q$.
 
\begin{figure}
\begin{center}
\resizebox{0.45\textwidth}{!}{
\includegraphics{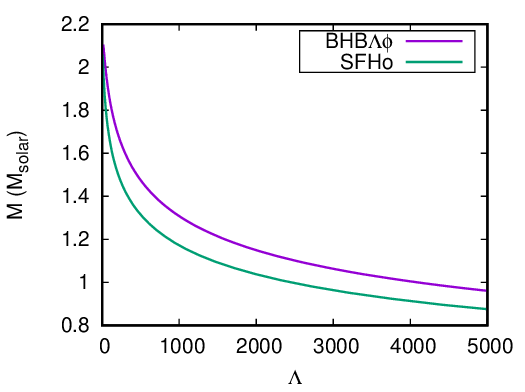}
}
\caption{\small{(Color online)
Mass of neutron star is plotted as a function of tidal
deformability for BHB$\Lambda\phi$ and SFHo equations of state.}}
\end{center}
\end{figure}

Neutron star mass is plotted as a function of individual tidal deformability 
for BHB$\Lambda\phi$ and SFHo EoSs in Fig. 4. The tidal deformability decreases
as the neutron star becomes more massive. This also results in higher 
compactness. Consequently, more compact neutron stars will be less deformed. 
The tidal deformability for a 1.4 M$_{\odot}$ neutron star is 697 and 334 in 
case of BHB$\Lambda\phi$ and SFHo EoSs, respectively.  

\begin{figure}
\vspace{0.7cm}
\begin{center}
\resizebox{0.45\textwidth}{!}{
\includegraphics{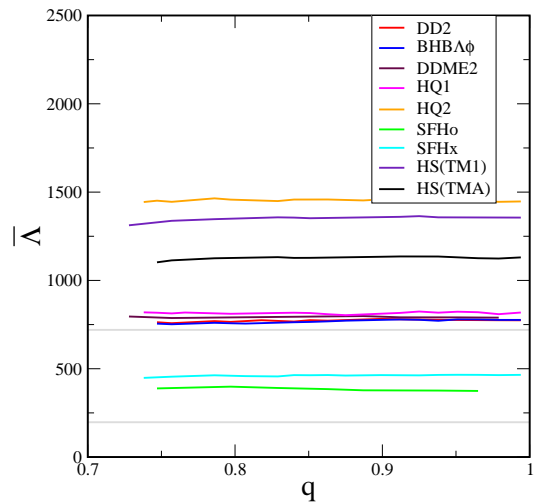}
}
\caption{\small{(Color online)
Tidal parameter $\bar{\Lambda}$ is plotted against mass ratio q for a fixed 
chirp mass ${\cal M}_{chirp}$ = 1.188 M$_{solar}$. Observational upper and lower
limits (grey lines) are shown here. }}
\end{center}
\end{figure}

LIGO and Virgo observations extracted the tidal contribution from the inspiral 
phase. The parameter ($\bar{\Lambda}$) that enters into the phase of the  
gravitational wave signal is a mass-weighted linear combination of individual 
dimensionless tidal deformabilities as \cite{fatava} 
\begin{eqnarray}
 \bar{\Lambda} = \frac{16}{13} \frac{(m_1 +12 m_2)m^{4}_1 \Lambda_{1} + 
(m_2 +12 m_1)m^{4}_{2} \Lambda_{2}}{(m_1+m_2)^{5}}
\end{eqnarray}
which is considered to be less than 720 at 90$\%$ confidence level 
\cite{ligo2} for low spin prior. Also an additional constraint placed on 
$\bar{\Lambda} \geq 197$ is based on EM observations of GW170817 
\cite{coughlin}. So the allowed window for $\bar{\Lambda}$ is now 
$197 \leq \bar{\Lambda} \leq 720$.

\begin{figure*}
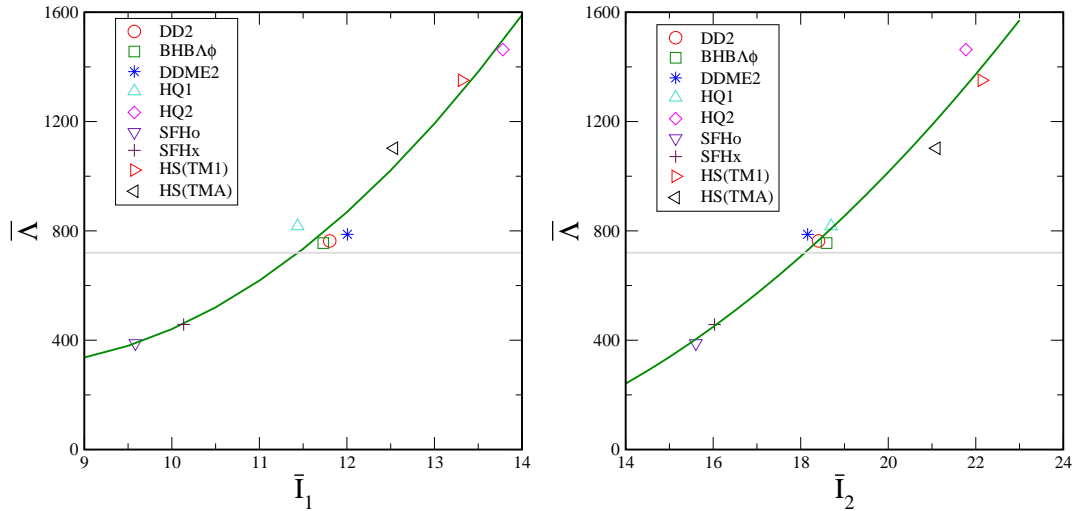

\vspace{0.8cm}
\begin{center}
\resizebox{0.45\textwidth}{!}{
\includegraphics{LI1.eps}
}
\resizebox{0.45\textwidth}{!}{
\includegraphics{LI2.eps}
}
\caption{\small{(Color online)
Mass weighted average tidal deformability parameter ${\bar{\Lambda}}$ is shown 
as a function of dimensionless moment of inertia of the heavier component of 
the neutron star binary having mass 1.58 M$_{\odot}$ (left panel) and 
and for the lighter component of the neutron star binary
having mass 1.18 M$_{\odot}$ (right panel).}}
\end{center}
\end{figure*}

Mass weighted average tidal deformability parameter $\bar{\Lambda}$ is plotted
with the ratio ($q$) of masses of merger components for a fixed chirp mass
${{\cal M}_{chirp}} = 1.188$ M$_{\odot}$ in case of low spin scenario in 
Fig. 5. Here 
results are shown for all EoSs. Upper and lower boundaries on $\bar{\Lambda}$
that were obtained from the gravitational wave and EM observations, 
respectively, are also shown in Fig. 5. It is found that results corresponding
to all EoSs satisfy the lower boundary. However, this can not be said about
several EoSs with respect to the upper boundary. It is evident from Fig. 5
that HS(DD2) and BHB$\Lambda\phi$ EoSs are marginally outside the upper boundary
whereas HS(TM1), HS(TMA), HQ1 and HQ2 EoSs are conclusively ruled out by the 
GW data. It is worth mentioning here that the estimates of both boundaries are
strongly model dependent \cite{radice,coughlin,ligo2}. It is also noted from
Fig. 5 that $\bar{\Lambda}$ is independent of $q$. This is also demonstrated 
analytically by Zhao and Lattimer \cite{zhao}.  

We calculate gross properties such as moment of inertia and quadrupole 
moment of slowly rotating neutron stars with spin frequency 100 Hz in this 
calculation using LORENE \cite{gourg,lorene}.
Figure 6 shows the relations between the parameter $\bar{\Lambda}$ versus
dimensionless moments of inertia of merger components ${\bar{I}}_{1}$ (left 
panel) and ${\bar{I}}_{2}$ (right panel), respectively. These results are 
obtained for masses of merger 
components ($m_1 =$1.58, $m_2 =$1.18) M$_{\odot}$ as obtained from 
the chirp mass. The upper bound on $\bar{\Lambda}$ at 90$\%$ confidence level
as obtained from gravitational wave data of GW170817 
is also included on the 
plot. The intersections of the curves with the upper bound of $\bar{\Lambda}$ 
give upper limits on 
the values of moments of inertia of two merger components. The values of 
${I}_1$ and ${I}_2$ so obtained are $\sim 2.0 \times 10^{45}$ 
g/cm$^2$ and $\sim 1.2 \times 10^{45}$ g/cm$^2$, respectively. These values of 
moments of inertia are consistent with the theoretically predicted values of 
Ref.\cite{sudip}. 
\begin{figure}
\vspace{0.7cm}
\begin{center}
\resizebox{0.45\textwidth}{!}{
\includegraphics{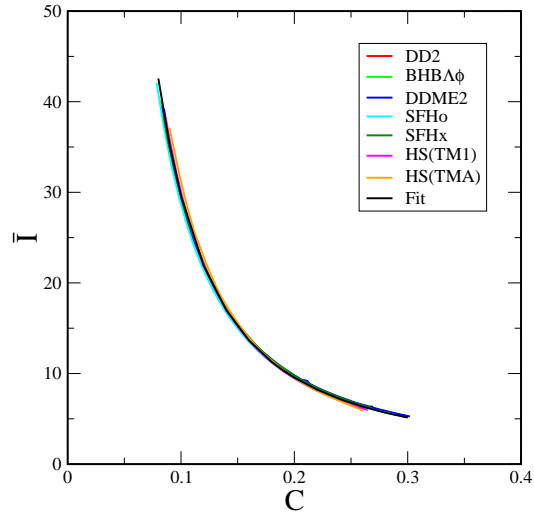}
}
\caption{\small{(Color online)
Dimensionless moment of inertia is plotted with compactness of neutron star 
for different hadronic equations of state.}}
\end{center}
\end{figure}

As we know the moment of inertia and mass of each component, 
it is possible to estimate the radius of the corresponding component. This is
done using the universal relation between dimensionless moment of inertia and
compactness of neutron star \cite{breu}. This universal relation is shown in
Fig. 7 for equations of state considered here except HQ EoSs in Fig. 7. The
universal relation is fitted with the functional form as given by Eq. (20) of 
Ref. \cite{breu}. It is evident that the upper limit on the tidal deformability 
constrains radii of merger components to be $\sim$ 13 km which are independent 
of component masses \cite{ozel}. It is worth mentioning here that HQ EoSs 
violate the universality \cite{deb}.  

We do the similar investigation for quadrupole moments ($Q_1$, $Q_2$) of merger
components in GW170817. Figure 8 exhibits the behaviour of mass weighted 
average tidal deformability parameter with dimensionless quadrupole moments 
${\bar{Q}}_1$ (left panel) and 
${\bar{Q}}_2$ (right panel), respectively. The upper limit on $\bar{\Lambda}$ 
from gravitational wave observation of GW170817 is also
shown in both figures. The values of upper bounds on quadrupole moments are  
found to be in the range 0.29 - 0.30 $\times 10^{43}$ g cm$^2$. Unlike the 
cases of moments of inertia in Fig. 6,
the estimated value of upper bound on $Q_1$ is less than that of $Q_2$
because the latter merger component is less compact and it is easy to deform 
the star. 

\begin{figure*}
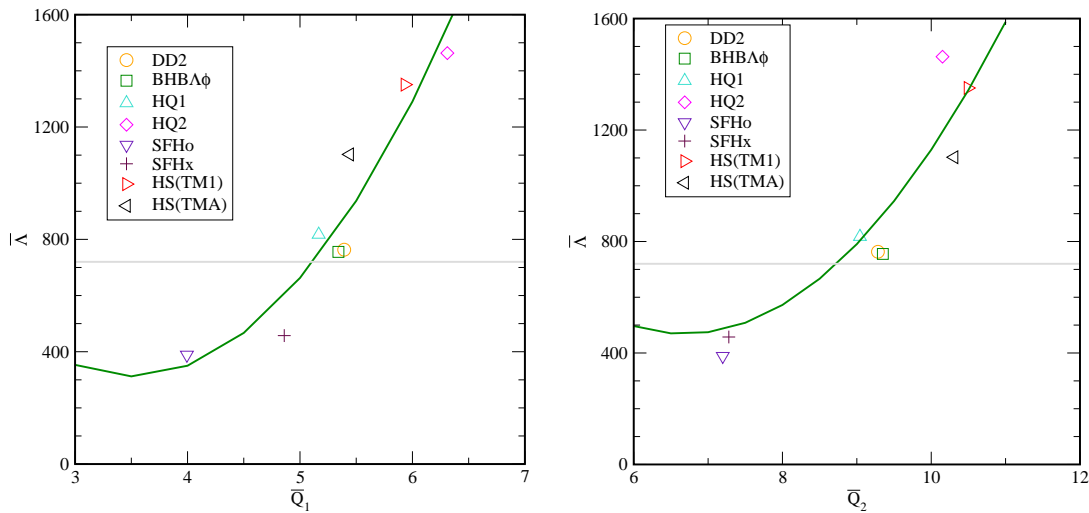

\vspace{1cm}
\begin{center}
\resizebox{0.45\textwidth}{!}{
\includegraphics{LQ1.eps}
}
\hspace{0.1cm}
\resizebox{0.45\textwidth}{!}{
\includegraphics{LQ2.eps}
}
\caption{\small{(Color online)
Mass weighted average tidal deformability parameter (${\bar{\Lambda}})$ is 
shown as a function of dimensionless quadrupole moment of the heavier component
(left panel) and lighter component (right panel).}}
\end{center}
\end{figure*}

\section{Conclusion}
We investigate the constraints on EoS of neutron star matter and properties of 
merger components of GW170817. We exploit large number of EoSs 
involving nucleons, hyperons and quarks in this study. We compute mass-radius 
relation, tidal deformabilities, moment of inertia and 
quadrupole moment of the merger components using the same EoS. All those 
quantities are dependent on EoS. We find that 50 $\%$ and
90 $\%$ credible intervals on the mass weighted average tidal deformability 
parameter ${\bar{\Lambda}}$ obtained from gravitational wave data of GW170817 
allow soft to moderately stiff EoSs. BHB$\Lambda\phi$ EoS might be allowed
by 90$\%$ credible interval whereas too stiff EoS like HS(TM1), HS(TMA) and 
HQ2 are ruled out. It is also observed that tidal deformabilities and mass 
ratio of merger components satisfy ${\Lambda_1}/{\Lambda_2} \sim q^6$ as 
predicted by other groups \cite{de,zhao}. Next we obtain upper bounds on 
moments of inertia and quadrupole moments of slowly rotating neutron stars 
exploiting the upper limit on the effective tidal deformability parameter 
${\bar{\Lambda}}$ of GW170817. It has been possible to estimate radii 
of two merger components $\sim$ 13 km as masses and moments of inertia of two 
merger components are known.

\end{document}